# Relative Grain Boundary Energies from Triple Junction Geometry: Limitations to Assuming the Herring Condition in Nanocrystalline Thin Films

Matthew J. Patrick[a], Gregory S. Rohrer[b], Ooraphan Chirayutthanasak[c], Sutatch Ratanaphan[d], Eric R. Homer[e], Gus L. W. Hart[f], Yekaterina Epshteyn[g], and Katayun Barmak[a,*]

[a]Department of Applied Physics and Applied Mathematics, Columbia University, New York, NY 10027, USA

[b]Department of Materials Science and Engineering, Carnegie-Mellon University, Pittsburgh, PA 15213, USA

[c]Department of Tool and Materials Engineering, Faculty of Engineering, King Mongkut's University of Technology Thonburi, Bangkok 10140, Thailand

[d]Department of Computer Engineering, Faculty of Engineering, King Mongkut's University of Technology Thonburi, Bangkok 10140, Thailand

[e]Department of Mechanical Engineering, Brigham Young University, Provo, UT, 84602, USA

[f]Department of Physics and Astronomy, Brigham Young University, Provo, UT, 84602, USA

[g]Department of Mathematics, University of Utah, Salt Lake City, UT 84112, USA

*Corresponding Author:
Katayun Barmak
kb2612k@columbia.edu
+1 (212) 854-8267
500 W 120th St, Suite 200
New York, NY 10027, USA

**Abstract**

Grain boundary character distributions (GBCD) are routinely measured from bulk microcrystalline samples by electron backscatter diffraction (EBSD) and serial sectioning can be used to reconstruct relative grain boundary energy distributions (GBED) based on the 3D geometry of triple lines, assuming that the Herring condition of force balance is satisfied. These GBEDs correlate to those predicted from molecular dynamics (MD); furthermore, the GBCD and GBED are found to be inversely correlated. For nanocrystalline thin films, orientation mapping via precession enhanced electron diffraction (PED) has proven effective in measuring the GBCD, but the GBED has not been extracted. Here, the established relative energy extraction technique is adapted to PED data from four sputter deposited samples: a 40 nm-thick tungsten film and a 100 nm aluminum film as-deposited, after 30 and after 150 minutes annealing at 400°C. These films have columnar grain structures, so serial sectioning is not required to determine boundary inclination. Excepting the most energetically anisotropic and highest population boundaries, i.e. aluminum Σ3 boundaries, the relative GBED extracted from these data do not correlate with energies calculated using MD nor do they inversely correlate with the experimentally determined GBCD for either the tungsten or aluminum films. Failure to reproduce predicted energetic trends implies that the conventional Herring equation cannot be applied to determine relative GBEDs and thus geometries at triple junctions in these films are not well described by this condition; additional geometric factors must contribute to determining triple junction geometry and boundary network structure in spatially constrained, polycrystalline materials.

1. Introduction

The grain structure of polycrystalline materials and the properties of their grain boundaries play a key role in determining electrical [1-3], mechanical [4], and chemical properties [5,6] in a variety of systems. The grain boundary character distribution (GBCD), which measures the relative areas of boundary types in a sample, has been shown to correlate with macroscopic properties in a variety of systems [7]. As an example, twin boundaries have been shown to contribute significantly less to resistivity than boundaries of different types [1-3]. To fully define the crystallography of a boundary, its character must be specified over 5 five macroscopic degrees of freedom; three of these parameters define the misorientation between the two neighboring crystallites, and two define the geometric plane which divides them. Even when accounting for symmetry in cubic systems, and dividing the space coarsely in increments of 10°, there are still more than 5,000 boundary types available, meaning large datasets of $>10^4$ boundary segments are usually required to obtain smooth distributions insensitive to the details of the sampling in this large space [8].

Advancements in automated electron backscatter diffraction (EBSD) have made the measurement of a large number of grains and grain boundaries routine in bulk microcrystalline samples. This has paved the way for several analysis techniques to be developed to determine grain boundary character and relative grain boundary energies. From a single layer of orientation data, four of the five degrees of freedom can be directly determined for any given boundary; these are the three parameters for misorientation and one of the two plane parameters. Applying stereological analysis to the traces of a large enough set of boundaries allows an accurate estimate to be made of the relative areas of boundary planes, and thus the five-dimensional (5D)

GBCD [8]. This stereological method has also been applied to nanocrystalline thin film data, collected using precession enhanced electron diffraction (PED) in the transmission electron microscope (TEM); furthermore, nanocrystalline thin film GBCD results in metals like copper [9], tungsten [10], and aluminum [11], have shown excellent correlation to comparable bulk materials.

To fully characterize the boundary network in bulk samples without the use of statistical approximations, techniques employing automated EBSD and serial-sectioning have been developed, allowing researchers to fully map the 3D boundary network, and directly measure the inclination of the boundary planes [12-14]. Methodologies taking advantage of this detailed 3D information have been developed to calculate relative grain boundary energies as a function of bicrystallography based on the geometry of triple junctions and an assumed force balance between the Hoffman-Cahn capillarity vectors [15] of each participating boundary [14,16,17]. First implemented on a magnesia sample [18], the complete 5D relative energy distribution has been calculated for a variety of systems. In each case, the boundary populations have been observed to have an inverse log-linear correlation with relative energy [13,19-22]. This inverse correlation has also been observed between boundary populations and absolute energies calculated from molecular dynamics (MD) and interpolation in both bulk [24,24] and thin film [25] samples, especially for high population and high energy-anisotropy boundaries.

While relative energy calculations have shown great success in bulk systems, the same analysis has yet to be conducted on thin-film PED data. For films like those discussed in this work, the grains are columnar [9-11, 26, 27]; with few exceptions, the grain boundary planes are perpendicular to the surfaces of the film. In tungsten films deposited as part of the same series of films as the one under study here, the observed grain size was independent of film thickness between 20 and 180 nm. For one of those tungsten films, which is significantly thicker than that under study in this work, cross sectional microscopy showed a distinctly columnar microstructure [27]. This unique structure can be exploited to greatly simplify the determination of the crystallographic plane of a given boundary segment, as it can be easily deduced from the orientation of the grains and the segment's trace in the crystal reference frame. As a result, the GBCD can be calculated from one layer of data without the use of stereology, yielding comparable results to the stereological methods; in both cases there is remarkable similarity to GBCD results from comparable bulk microcrystalline materials [9-11]. It would therefore appear possible to perform the described relative energy extraction with just one layer of PED orientation data from nanocrystalline films, under the assumption that the triple lines are perpendicular to the plane of the film and run through the film thickness from the top surface to the substrate interface. Some prior work suggests that a single layer of data is sufficient in films. A grain boundary energy reconstruction method was applied to a single layer of EBSD data from a 1.7 μm-thick, microcrystalline, aluminum film to reconstruct the relative energies of [111]-tilt boundaries as a function of misorientation angle; in that work, the film was assumed to have a columnar microstructure, with boundaries perpendicular to the film surface at least as deep as the electron escape depth, and it was assumed that the dihedral angles of the considered triple junctions was solely determined by the relative energies of the boundaries [28].

The purpose of this work is to test whether the Herring condition provides an appropriate general description of the equilibrium at triple junctions in thin films and whether it can be used to reconstruct the relative GBED. This has already been successfully demonstrated in bulk systems [19-22], however the role of thin film geometry is as yet unknown. The significance of this aspect of the microstructure is emphasized by the results of this research. To approach this problem, the most recent method for the calculation of relative energies from triple junction geometry [17] is adapted to thin film PED data of four samples. The relative GBED results are then compared to their corresponding GBCDs and to grain boundary energies from MD calculations and interpolation. The expected energetic trends are not recovered, ultimately calling into question whether the conventional Herring equation accurately describes the geometry of grain boundary triple lines in thin films where the grain size is larger than the film thickness. This insight points to other driving forces in the development of the geometry of the grain boundary network and character distribution in thin films and is consistent with the results of recently developed mathematical theory and simulations modeling grain growth with dynamic lattice misorientations and with finite triple junction mobility [29,30]. These findings have implications for future grain growth simulations of nanocrystalline materials and experimental design to more effectively study these technologically relevant materials.

## 2. Methods

### 2.1: Sample and Data Preparation

In this study, two data sets are examined. First, data from a nominally 40 nm-thick α-tungsten film is analyzed. The film was sputter deposited on an oxidized Si(100) substrate and encapsulated with an underlayer and an overlayer of sputtered silicon dioxide in order to provide identical top and bottom electron scattering surfaces. The film was subsequently annealed at 850°C for 2 hours to transform all β-W to α-W and was mapped using PED with a 0.3° precession angle and a step size of 5 nm. Film preparation, characterization, and 5-dimensional GBCD have been described in detail elsewhere [10]. Then, a 100 nm-thick aluminum film, also sputter deposited on thermally oxidized silicon, is analyzed in its as-deposited state as well as after 30 minutes and 150 minutes annealing at 400°C; these films and their characterization were also previously described in detail [11] and were mapped using a precession angle of 0.6° and a step size of 4 nm for the as-deposited film and 5 nm for the annealed films. In contrast to the W film, the Al film was deposited directly onto an oxidized Si(100) substrate and was protected only by its native oxide on the top surface. The Si/SiO$_2$ substrates were chemically back etched to electron transparency using a nitric acid and hydrofluoric acid solution, similar to what has been described in detail elsewhere [31,32].

For each sample, the PED data were recorded and indexed using the ASTAR™ (NanoMEGAS, Brussels, Belgium) system and imported into the TSL OIM™ 8.1 (EDAX, Mahwah, NJ, USA) software package, where they were subject to a cleanup procedure. First, the grains, defined as a group of neighboring pixels with a maximum of 5° disorientation, were dilated to eliminate poorly indexed points which create spurious grains and boundaries. This was done with a minimum grain size of 5% of the mean area (cluster of 7 steps for Al, cluster of 15 steps for W), determined from the reconstructed boundary network prior to cleanup. A single average orientation was then assigned to each grain. Finally, the data were subject to the pseudo-symmetry cleanup, removing false boundaries that arise when two symmetry related orientations are assigned within a single grain, because the diffraction pattern is aligned with a rotational symmetry axis. This step is described in more detail in [33]. The aluminum and tungsten films both had false boundaries with 180° misorientation removed with a 1° tolerance angle for 20 axes. Tungsten was further subject to a cleanup of boundaries with misorientation of 60° about [111], with a tolerance angle of 2°. The final Euler angles and spatial coordinates associated with each boundary segment were exported, deviating from the true boundary position by no more than 2 steps. Generally, the GBCD is insensitive to the cleanup, but the reconstructed boundary network contains many fewer unphysical features. Further details on the cleanup and its impacts can be found in the Supplemental Information.

**2.2: Relative Energy Reconstruction**

To calculate relative energies, the geometry of the boundary network at the intersections of boundary planes is considered, and their capillarity vectors are calculated based on the assumption of the Herring condition of force balance. To accomplish this, triple junctions in each layer are identified as locations where the endpoints of the traces of three boundary segments meet. By comparing the locations of triple junctions in a given layer to the next layer, the direction of the triple line vector can be determined. The locally optimal block preconditioned conjugate gradient (LOBPCG) method is then implemented to solve the system of equations by minimizing the residual differences between capillarity vectors of boundaries and those of their neighbors in the 5D boundary space, while satisfying the Herring condition; then, by multiplying the resulting capillarity vector by the plane normals, relative energies for all boundaries within the sampled region can be calculated [15,17]. This nonparametric approach does not rely on the discretization of the 5D boundary space as the conventional method does [16,17] and therefore does not require uniform sampling of the 5D space to be successfully implemented. For this reason, the method is insensitive to orientation texture and can be used to calculate relative energies from data-sets with boundaries from limited regions of the 5D space, like the annealed aluminum films considered here. In the work which introduced this technique, for most boundaries, the uncertainty in the resultant relative energies was measured to be less than ±0.1, and had general errors of ±0.02 for regions outside cusps and ±0.04 for regions near cusps in the GBED [17].

To calculate relative energies from a single layer of PED data, the grain boundaries are treated as two identical sets of segments which are separated by the thickness of the film; triple junctions are identified and triple lines are constructed through the thickness of the film with the sample normal direction, $[001]_s$, where the subscript denotes sample directions. In total, there were 24,192 (57,623) triple junctions (total segments) identified in the W dataset, 29,723 (70,610) in the as-deposited Al film, 31,533 (67,356) in the Al film annealed for 30 minutes, and 17,511 (41,450) in the Al film annealed for 150 minutes. For certain boundaries, the plane is crystallographically constrained to a certain set of indices. As a relevant example, Σ3 boundaries with a trace in the [111] zone are overwhelmingly coherent twin boundaries, and so their planes are known to be (111)-type. In the aluminum samples, boundaries are identified as likely coherent twins by trace analysis, with a 10° tolerance. Then, the normals of the twins and the normals of the boundaries which meet with them at triple junctions are adjusted such that the triple line lies in the (111) boundary plane of the coherent twin rather than normal to the film surface. In the as-deposited (annealed) Al films, 7.5% (6.5%) of all triple lines were identified and adjusted. From this point, the energy reconstruction proceeds as described in [17] without further modification.

## 3. Results and Discussion

### 3.1 Tungsten Film

Figure 1 shows a representative inverse pole figure map with its corresponding inverse pole figure after cleanup of the 40 nm-thick tungsten film. Here, it is clear that the orientation texture is weak, with a maximum intensity of less than 1.5 MRD. The grains have an average size of about 100 nm, but experienced no grain growth, owing to the low homologous temperature of annealing. When the boundaries are reconstructed (shown as solid black lines in Figure 1a), the disorientations across the segments closely match the distribution for randomly misoriented cubes [34]. Figure 2 plots the disorientation distribution for the tungsten film as a solid black line and the distribution for the random case is plotted as a dotted green line. This close correlation indicates that the sample has no preference for any given misorientation angle.

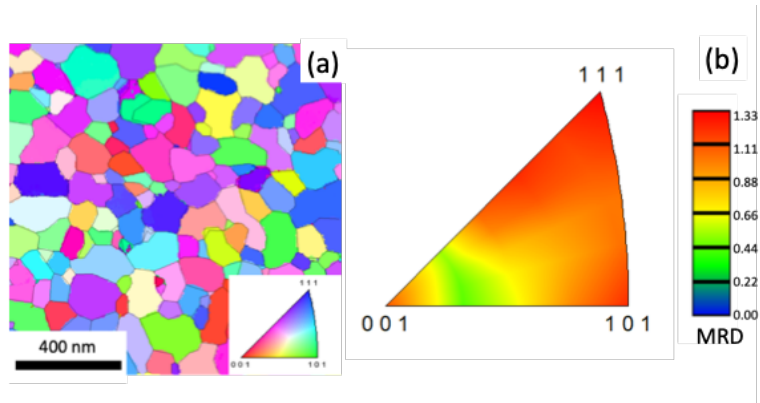

Fig. 1. (a) Inverse pole figure map along the sample normal, showing orientations of grains in a representative field of view of the nominally 40 nm-thick tungsten film. Reconstructed boundary segments are drawn as thin black lines. (b) Inverse pole figure, along the sample normal, for this field of view; it shows very weak orientation texture, with a maximum of 1.33 multiples of random distribution at the [111] direction.

Despite the lack of grain growth or clear preference for disorientation angle, the grain boundary plane distribution (GBPD) at fixed misorientations, three of which are shown in Figures 3a-c, is non-random and correlates well with bcc materials with grain size in the micrometer scale [10]. The populations of boundaries similarly have a clear qualitative inverse correlation with energies calculated via MD simulations and an interpolation scheme, as previously reported by [25,35]; corresponding GBEDs calculated by Chirayutthanasak et al. [25] in this manner are shown in Figures 3g-i. As expected, the logarithm of the population of boundaries showed a clear linear inverse correlation with the calculated energies [25], indicating that the lowest energy boundaries appear more frequently than higher energy boundaries.

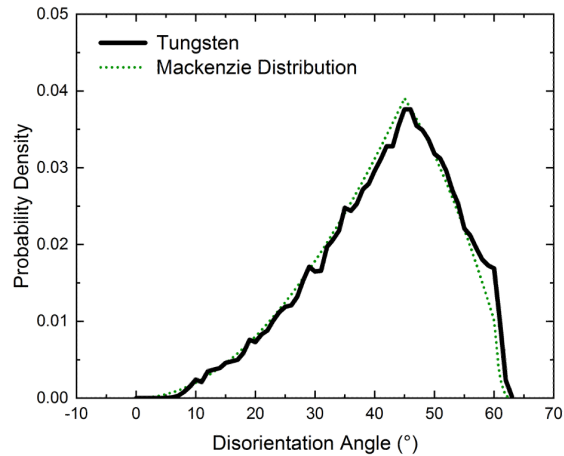

Fig. 2. Probability density of boundary disorientation angle across all reconstructed boundaries in all fields of view for the 40 nm-thick tungsten film plotted as a solid black line and the random, i.e. Mackenzie, distribution for disorientation [34] plotted as a dotted green line.

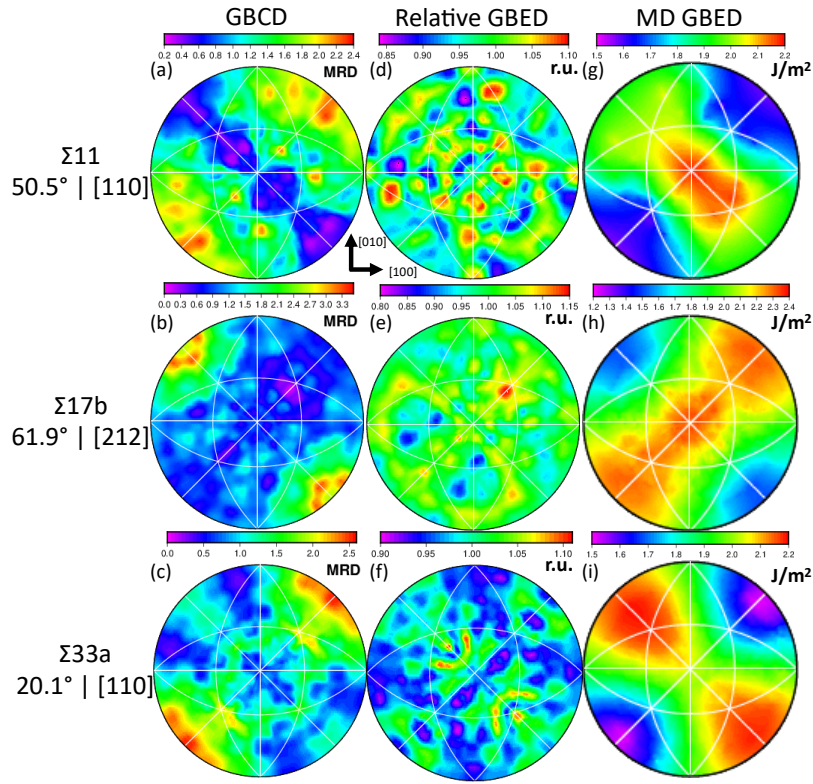

Fig. 3. Grain boundary plane distributions for the 40 nm-thick tungsten film, calculated using the stereological method [3] for Σ11, Σ17b, and Σ33a misorientations (a-c). There are 1,161, 955, and 1,195 boundary segments represented in each figure respectively. The corresponding relative energy distributions for the same set of boundaries are shown in the second column (d-f). Here, r.u. stands for relative units, and is referenced to the average relative boundary energy. Finally, in the third column (g-i), the energies calculated based on the interpolation scheme presented in [25] are shown.

When the described energy reconstruction method is applied to the data collected from the under the assumption that the Herring condition of force balance is satisfied, the LOGBC algorithm rapidly converges, with negligible residuals and just four segments returning invalid (negative) energies; the resulting GBED, however, fails to reproduce the energetic trends found in MD calculations and interpolation from [25] and fails to show a qualitative inverse correlation with the GBCDs plotted at fixed misorientations (c.f. Figures 3d-f). When the boundary populations are binned against their energies for the misorientations shown, as they are in Figure 4, there is no obvious relationship between the relative energy and the boundary populations; values of 1 to 2 MRD across the range of relative energies indicate that the lowest energy boundaries appear with similar area fraction to the boundaries with the highest relative energies. When a linear least-squares fit is applied to the data, for the Σ3, Σ17b, and Σ33a, the slope of the lines is not significantly different from zero when considering the standard error. For the Σ11 boundaries, the fit obtained has a negative slope, with an *R* value of -0.68. However, comparison between the stereographic plots shown in Figure 3 indicated the trend is likely spurious.

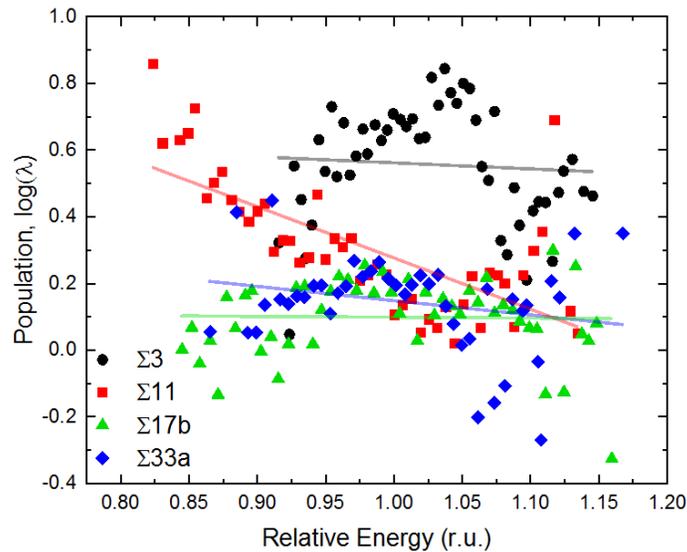

Fig. 4. Logarithm of boundary populations, log(λ) with λ in units of MRD, binned by their relative energy for the 40 nm-thick tungsten film for misorientations of Σ3 (black circles), Σ11 (red squares), Σ17b (green triangles), and Σ33a (blue diamonds). Least squares fit lines are plotted in the same color as their corresponding misorientation.

The failure of these calculations to reproduce the anticipated features and trends in the GBED indicates that the application of the Herring condition in this film is not appropriate for relative energy extraction in this sample. As the grain boundary energy is independent of microstructure, this result further implies that the geometry of the triple junctions is not fully determined by the conventional Herring force balance as was assumed during the calculations. It is worth noting, however, that this sample experienced no grain growth during annealing as the annealing temperature of 1120 K was less than 1/3 of the melting point of tungsten. This sample therefore inherits its structure from its deposition conditions and the nucleation rate of α-W in the originally deposited β-W grains [10]; it did not experience any major rearrangement of its grain boundary network in the course of annealing, which may suggest that the observed boundary-network geometry does not reflect an equilibrium state for boundary segments meeting at the triple junctions.

**3.2: Aluminum Films**

To address the limitations of the tungsten sample, the same analysis was conducted on the 100 nm-thick aluminum films, which were characterized in an as-deposited state as well as after being annealed for 30 minutes and 150 minutes at 400°C. In contrast to the tungsten film, the aluminum sample experienced significant grain growth and orientation texture evolution during annealing. The starting equivalent circle diameter of mean grain area was 109 nm, grew to 152 nm after 30 minutes, and grew to 157 nm after 150 minutes. The orientation texture evolved from a 2.4 MRD to 5.4(6.0) MRD preference for the [111] orientation after being annealed for 30(150) minutes [11]. Figures 5a-c shows representative inverse pole-figure maps from each annealing condition. With the exception of a notable excess at 60° and the corresponding deficit near the middle of the distribution, the disorientation distributions for all three films closely follow the Mackenzie distribution, as shown in Figure 6. The distribution of disorientations in this work has fewer departures from the random distribution than the previously reported distribution from the same sample [11], reflecting the cleanup procedure used here which includes the removal of pseudosymmetry boundaries to ensure a maximum number of usable triple junctions (see the Supplemental Information for more details.) This indicates that there is little preference for any given misorientation angle, except of 60°, corresponding to the Σ3 boundaries which are present in unusually high fractions in these three samples [11].

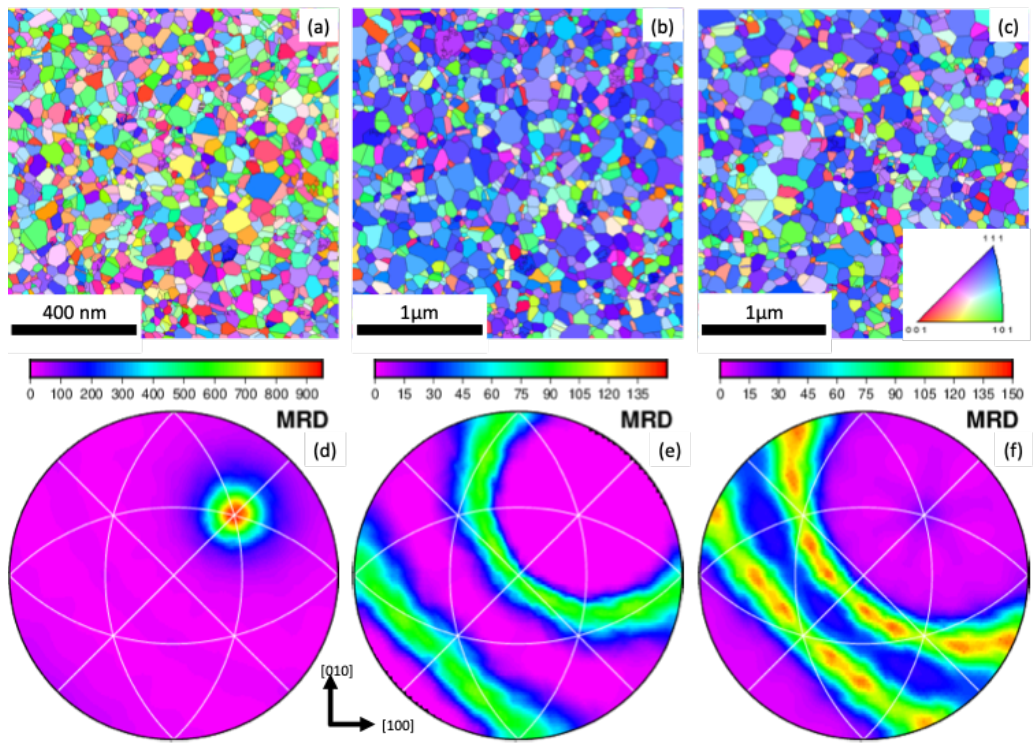

Fig. 5. Representative inverse pole figure maps for the 100 nm-thick aluminum films in their (a) as-deposited state and the films after annealing at 400°C for (b) 30 minutes and (c) 150 minutes. The corresponding grain boundary plane distributions at a fixed misorientation of 60°|[111] (after twin adjustment) are shown below their maps in (c-f).

After annealing, the orientation maps show an obvious preference for the [111] fiber texture. The GBPD, plotted at a fixed Σ3 misorientation of 60°|[111] (after twin adjustment), shown in Figures 5d-f, show a clear evolution in the boundary texture. The plane distribution for the Σ3

misorientation in the as-deposited film has a large peak of >900 MRD at (111), the pure twist location. The as-deposited sample has very little orientation texture, with a slight preference of 2.4 MRD for the [111] orientation [11]. The Σ3 twist boundaries are virtually eliminated and the plane distribution shows an increasingly strong preference for tilt boundaries, as can be seen most clearly in Figure 5f where the majority of Σ3 boundaries lie near the great circle 90° away from the (111) position. This reflects the increasingly limited available characters for boundaries as the film develops stronger fiber texture, since, if the boundaries are perpendicular to the film surface, two adjacent [111] oriented grains are constrained to meet at [111]-tilt boundaries.

Like the tungsten film, these samples have a mostly columnar structure. Indeed, the GBCD is insensitive to whether the segments are treated using stereology to determine relative areas as a function of boundary plane or treated as strictly columnar and determining boundary plane based solely on the boundary's trace direction and crystalline orientation. Direct comparisons can be found in the Supplemental Information. One exception to this are the low-energy coherent twin boundaries, however; these boundaries are not suitable for comparison because they were assumed to have the (111) orientation in the data processing.

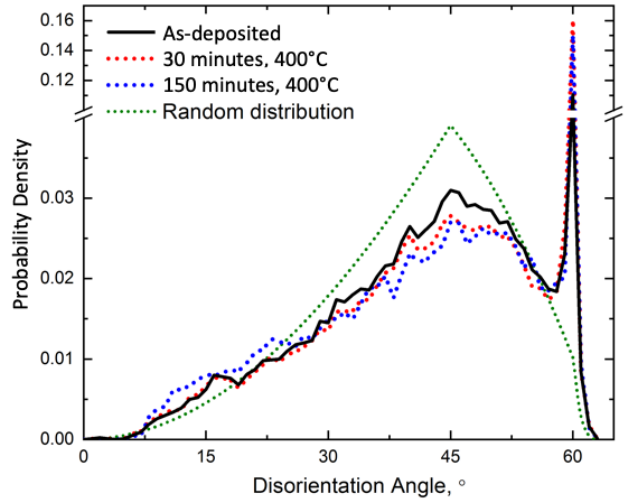

Fig. 6. Disorientation distribution for the 100 nm-thick aluminum films, reflecting the cleanup procedure used in the current work, plotted in solid black (as-deposited), red (30 minutes, 400°C), and blue (150 minutes, 400°C). The Mackenzie distribution for disorientations of randomly orientated cubes is shown as a dotted green line.

When the energy reconstruction is applied to the Al data-sets treated as fully columnar (before the twin adjustment), as it was in the case of tungsten, no relationship between population and the extracted relative energy is observed for any boundary types. After adjusting the twin boundaries and their triple lines, however, the GBED becomes consistent with expectations for Σ3 boundaries, but not other misorientations. Figures 7 and 8 contains twin-adjusted grain boundary plane distributions calculated using stereology and the relative GBEDs at selected misorientations of Σ3, Σ7, and Σ11 for the as-deposited film and the film annealed for 150 minutes at 400°C, respectively. The results for the 30 minute film may be found in the Supplemental Information, but closely match the results presented for the other two Al films. In spite of drastic differences in the grain boundary textures, the relative energy results from the annealed films and the as-deposited film are remarkably similar, indicating that the energy reconstruction is able to identify common relationships between triple junction geometry and their crystallography. Further, this consistency provides concurrent evidence that the columnarity of the as-deposited film is not significantly different from the annealed films; this indicates that the starting structure is largely columnar to begin with, validating one of the fundamental assumptions of the technique under study. In cases of very high energy anisotropy like misorientations about [111], it is able to reproduce distributions which approach the expected form determined via MD calculations. Even in cases where the energy reconstruction results do not match the predictions at all, as is the case of the Σ11 boundaries, the results are similar between the three films. In this way, the relative energy results successfully reflect the fact that the GBED is an intrinsic material property, independent of the microstructure. In both presented films, at the highest population misorientation (60°|[111], Σ3), Figure 7d (8d) show a deep minimum at the coherent twin position of (111), and a band of generally higher relative energy and a corresponding deficit in population is seen along the (111) tilt boundaries, as is predicted for fcc materials. Furthermore, the calculated relative energies show general agreement with the energies calculated via MD [36,37], and plotted in 7g (8g).

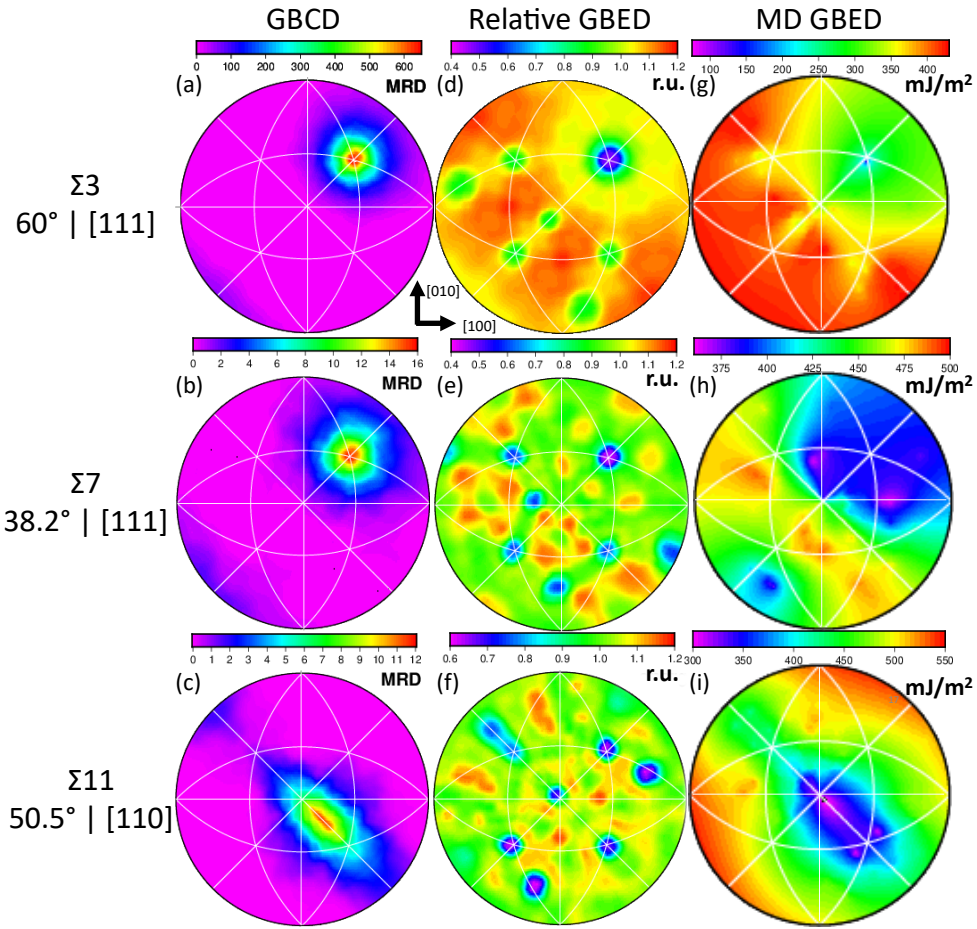

Fig. 7. Grain boundary plane distributions for the as-deposited aluminum film calculated using the stereological method and plotted in stereographic projection along the [001] direction for fixed misorientations of Σ3, Σ7, and Σ11 (a-c). There are 4,491, 1,585, and 2,020 boundary segments represented in each figure respectively. The corresponding relative energy distributions are shown for the same set of boundaries in the second column (d-f). Here, r.u. stands for relative units, referenced to the average relative boundary energy. Finally, the absolute energies computed via molecular dynamics simulations are plotted in the third column (g-i). Note the minima at the (111) twist position for the Σ3 and Σ7 misorientations, with minimum values of 75 mJ/m$^2$ and 271 mJ/m$^2$ respectively, which are vanishingly narrow in the plots due to the limited sampling of the 5D space by the MD calculations [36,37]. Further, the appearance of the Σ7 symmetry is affected by the contouring between sparse points during plotting.

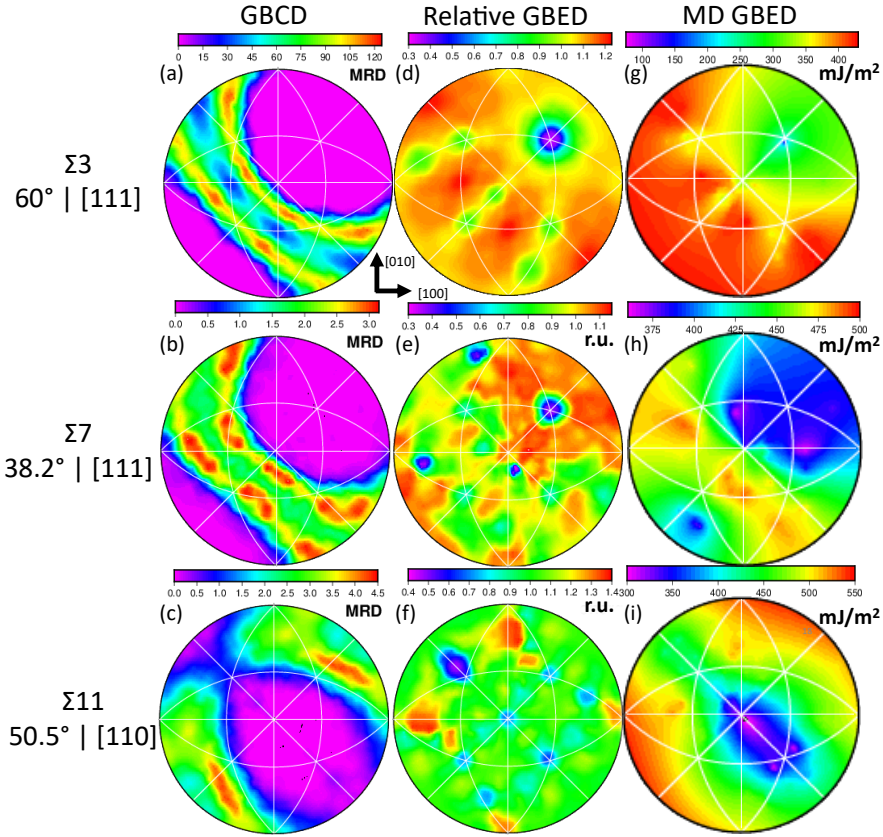

Fig. 8. Grain boundary plane distributions for the aluminum film annealed at 400°C for 150 minutes calculated using the stereological method and plotted in stereographic projection along the [001] direction for fixed misorientations of Σ3, Σ7, and Σ11 (a-c). There are 2,486, 760, and 1,158 boundary segments represented in each figure respectively. The corresponding relative energy distributions are shown for the same set of boundaries in the second column (d-f). Here, r.u. stands for relative units, and are referenced to the average relative boundary energy. Finally, the absolute energies computed via molecular dynamics simulations are plotted in the third column (g-i). Note the minima at the (111) twist position for the Σ3 and Σ7 misorientations, with minimum values of 75 mJ/m$^2$ and 271 mJ/m$^2$ respectively, which are vanishingly narrow in the plots due to the limited sampling of the 5D space by the MD calculations [36,37]. Further, the appearance of the Σ7 symmetry is affected by the contouring between sparse points during plotting.

When other misorientation angles about the [111] direction are considered, like Σ7 boundaries (38.2°|[111]), the results for the as-deposited film (Figure 7b,e) still show partial qualitative agreement between the populations and relative energies; the annealed films do not, however, due to their orientation texture. For both the as-deposited and the annealed film, Figures 7e (8e) show that the reconstruction again captures the predicted energy minimum at the Σ7 twist position of (111), and the band of high energies along the (111) tilt boundaries, which overlaps with the maxima found via MD calculations (Figures 7h, 8h). This is consistent with the observed GBCD for the untextured as-deposited film, although the quantitative correlation between the populations and the extracted relative energy when plotted and fitted (Figure 9) is weak, with an $R$ value of -0.43. Furthermore, there are sharp minima present in the Al Σ7 relative GBEDs at arbitrary locations, which neither correspond to features in the GBCDs, nor to features found in bulk fcc materials' relative energy functions (e.g. nickel [21]), nor to features calculated by MD [36,37]. Finally, for boundaries with misorientation axes other than [111], the extracted relative energies shows virtually no correlation with GBCD, correlation with the GBEDs of other fcc materials [21], nor MD calculated energies, as can be seen qualitatively in Figures 7c,f, and i and 8c,f, and i, representing distributions for the Σ11 boundaries. Here, both of the experimental energy distributions have a minimum at the $(\bar{3}32)$ symmetric tilt grain boundary (STGB) while the MD calculated energy distribution has a minimum at the $(1\bar{1}3)$ STGB and the experimental GBCD has a maximum at the $(1\bar{1}3)$ STGB.

In Figure 9, where the Al boundary populations are binned against their relative energies, the scatters follow no consistent trends. For all but the Σ3 boundaries, the distributions are relatively flat. When a linear least square fit is applied to the as-deposited data, the Σ7 boundary populations have a negative slope, but have an *R* value of just -0.43; Σ11 boundaries' population have a slightly more linear trend, with an *R* value of 0.69, but they show a weakly *positive* correlation with relative energy; Σ19a boundaries have a slope which is not significantly different (given standard errors) from zero. For the 150 minute film, Σ7 boundaries show a positive correlation with relative energy, in contrast to a much stronger negative correlation (*R* = -0.75) found between the same populations and the MD energies, which is plotted in Figure 10; Σ11 and Σ19a boundaries do not have slopes significantly different from zero.

The Σ3 boundaries do show some inverse correlation between population and energy over a small range of energies in both films. This correlation breaks down for boundaries with relative energies above ~0.8 r.u., yielding two distinct regimes. For the as-deposited (150 minute) film, the *R* value is -0.75 (-0.99) at energies below the discontinuity and -0.64 (-0.34) at energies above it. This disjointed result is especially evident in the annealed sample, where the high population of the relatively high energy tilt boundaries is observed because of the film geometry. In contrast, when the boundary populations from the as-deposited film are plotted against energies calculated by molecular dynamics [36,37], as in Figure 10, the inverse correlation shows no such discontinuities for Σ3 boundaries, or any of the other selected boundary misorientations.

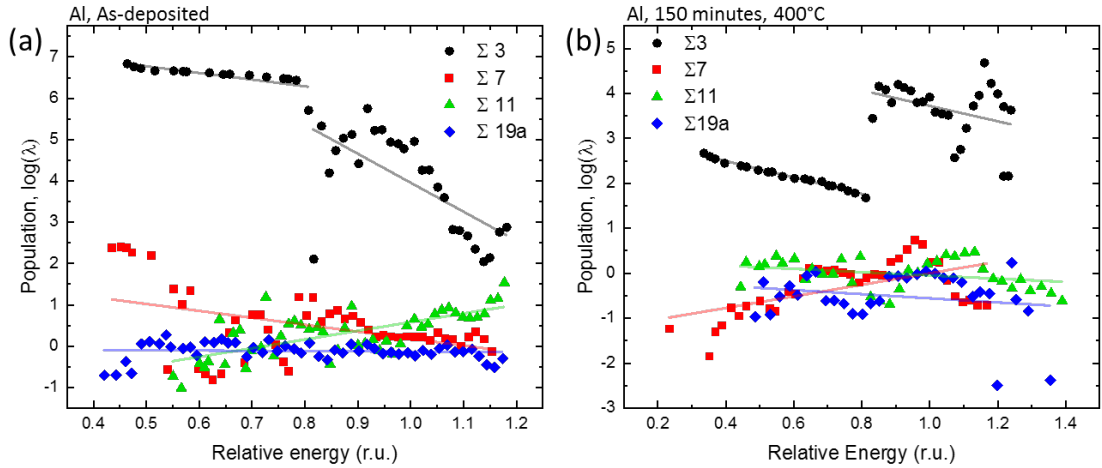

Fig. 9. Logarithm of boundary populations, log(λ) with λ as relative area in units of MRD, as binned by their relative energy for the aluminum film at Σ3, Σ7, Σ11, and Σ19a misorientations for the (a) as-deposited film and (b) the film annealed for 150 minutes at 400°C. Least squares fit lines are plotted in the same color as their corresponding misorientation, including two regimes for the Σ3 boundaries.

While the relative energies extracted under the assumption of local equilibrium at the junctions do not reproduce trends observed in other systems, when the boundary energies computed by Homer et al. [36,37] are compared to nanocrystalline experimental populations, the inverse relationship is observed. Figure 10 shows the population values, read from the GBCD, plotted against the corresponding energy for each boundary considered by Homer et al. [36,37]. For the entire range of boundaries, shown in Figure 10a and 10c, grain boundaries with the lowest energy have the highest populations, and those with the highest energies tend to have lower populations. For boundaries which do not appear frequently in the experimental dataset or have low anisotropy, this correlation is generally weak. The as-deposited Σ11 GBCD, for example, clearly matches the expectations set out by the MD energy results presented in Figure 8, but the quantitative correlation observed when fit is very weak, with an |R| of < 0.1. For misorientations

with very high populations (>100 MRD for at least one plane normal), however, a clear log-linear inverse correlation is observed, with *R* values of -0.95 for the Σ61d misorientation, -0.92 for the Σ3 misorientation, -0.74 for the Σ131e misorientation, and -0.62 for the Σ67d misorientation. For those plotted, all four have (or are close to) the [111] disorientation axis, and are generally close to the Σ3 boundaries, explaining their generally high population (Σ3: 60°|[111], Σ61d: 52.7°|[111], Σ67d: 60.5°|[443], Σ131e: 60.3°|[554]). For misorientations whose

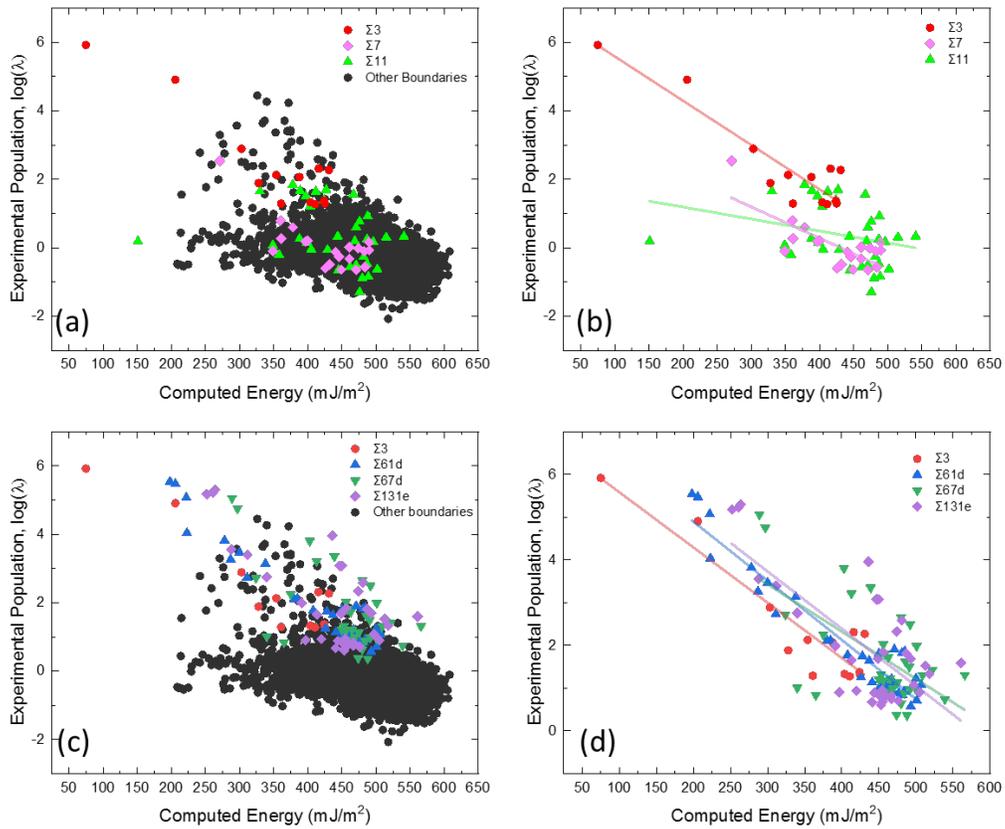

Fig. 10. Logarithm of boundary populations, log(λ) with λ as relative area in units of MRD, in the as-deposited aluminum film plotted against boundary energies computed by molecular dynamics in [36,37]. All 7,304 examined boundaries are shown in (a) and (c), with the boundaries with misorientations specifically discussed in this work highlighted in color in (a) and (b), and boundary types with at least one boundary type with population >100 MRD highlighted in (c) and (d). Least squares fit lines are plotted in the same color as their corresponding misorientation.

GBPD are shown in Figure 7, especially the Σ3 and Σ7 boundaries, the relationship is also recovered, as one might expect comparing Figure 7(a-c) to Figure 7(g-i), with $R$ values of -0.92 and -0.75 respectively.

Finally, consistent with the analysis above, the energies computed under the assumption of the conventional Herring force balance and those computed using MD show no correlation to one another. Figure 11 plots the extracted relative energies for the 7,304 boundary types considered in [36,37] against the energies computed by MD in [36,37]. Here, the trend has a correlation coefficient of < 0.1. Taken all together, these results reinforce the evidence that the characters taken by the boundaries in the aluminum films correspond to configurations which lower the total system energy, but the local equilibrium at the triple junctions is governed by factors not described by the Herring equation.

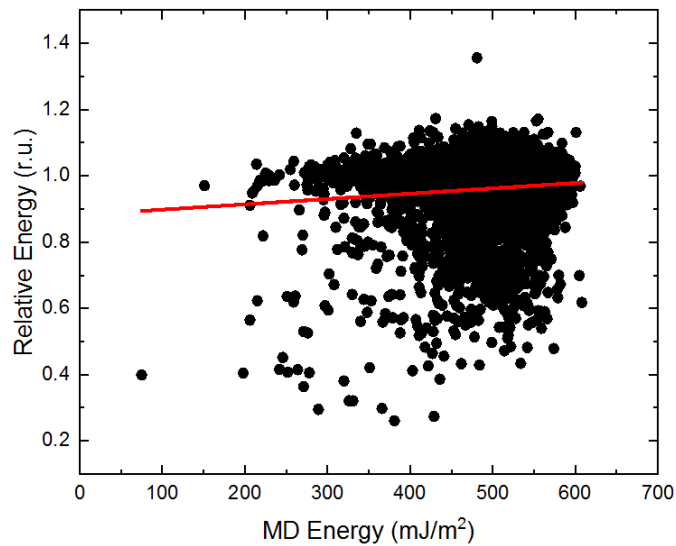

Fig. 11. Relative energies extracted from triple junction geometry of the as-deposited film for the 7,304 boundary types examined in [37,37], plotted against the grain boundary energies calculated via MD in [36,37]. The least squares best fit line is plotted in red, with a correlation coefficient < 0.1.

## 3.3: Discussion

The grain boundary energy distribution measured from thin films with a variety of annealing conditions, orientation texture, grain size, and crystal structures is inconsistent with expectations. The inverse correlation between energies and populations is not observed and the extracted relative energies are uncorrelated with MD calculated energies. This departure indicates that the Herring condition for local equilibrium cannot be applied to extract relative energies in thin films as it can for bulk microcrystalline samples, and thus, in contrast to bulk samples, the triple junction geometry in thin films is not well-described by the Herring equation. In these thin films, other effects which are not dominant in those bulk systems and not factored into the Herring equation must play a significant role in determining the behavior of the network, like residual stresses, free-surface energies, and geometric constraints. To successfully describe the behavior of boundaries at triple junctions in films like those under study here, the force balance at the junctions will have to include some factors beyond the simplest relationship between grain boundary energy and geometry, as the evidence presented here suggests that the local equilibrium condition is more complex than previously understood.

As an example, significant stresses develop during vapor deposition of thin films [38], and large biaxial thermal stresses remain after annealing [39]. Due to mechanical anisotropy in many materials, certain grain orientations reduce volumetric strain energy; the minimization of this energy is a documented driving force for grain growth [38,40], and consequently of grain boundary migration, meaning that this factor applies forces to boundaries which are not considered in the conventional Herring equation's description of local equilibrium. Some work has already been done to locally measure strain gradients and estimate dislocation density with PED [41,42], which could be applied to assess the contributions of strain energy to the behavior or TJs in thin films. Surface energy minimization is another well documented driver of grain growth [40,43], and so surface energies must similarly be responsible for some forces applied to the boundaries, presumably influencing triple junction geometry. A purely thermodynamic force is also present: because the minimum-area boundary is achieved when the boundary is perpendicular to the film plane, rotations away from this orientation are energetically costly compared to rotations around the film axis, but neither these rotations nor the change in area are considered in the Herring equation. Each of these effects would exert forces not considered by the conventional Herring equilibrium equation and could contribute to the erroneous relative energies extracted by the technique used in this work.

Despite these complications, correlation between MD calculated energies and populations, as well as the consistently observed correlations between measured bulk and thin-film GBCDs, reveal that geometric constraints present in thin film are only exerting strong influence on the triple junction geometry, not on the selection of boundary character; the GBCDs still closely follow the thermodynamic expectations as they do in bulk materials. In other words, the triple junction geometry is not determined solely by the energies of the participating grain boundaries; thus, accurate relative grain boundary energies cannot be extracted by only considering the geometry of triple lines through the thickness of the film. In contrast, for samples with randomly oriented grains, the grain boundary character distributions are overwhelmingly determined by the grain boundary energies. In this work, because the relative energies were extracted based on the conventional Herring equation, which links triple junction geometry exclusively to energies of the grain boundary segments in the neighborhood of the junction, they do not reflect the true energies and so they are uncorrelated to the boundary character distributions and the energies computed via MD.

The strong correlation between populations and computed energies, however, indicates that boundary populations obtained from thin-film experiments can be used to qualitatively assess relative boundary energies and be used to validate computed boundary energies via their populations, as is suggested by Holm et al. with respect to bulk materials [23]. Starting from MD calculations of pure materials, studying changes in GBCD as a function of impurity content or alloying additions, for example, would allow the study of their relative effect on grain boundary energies. Furthermore, experiments must be conducted to determine under what conditions, if any, local triple junction equilibrium can be accurately described by the conventional Herring equation in thin films. In conjunction with calculated energies and the GBCD, the relative grain boundary energy extraction as performed in this work can act as a proxy to test whether the conventional equilibrium condition is met in any given system.

To isolate the effects of surface energy, for example, the analysis presented here can be repeated on highly [111]-textured samples, where every grain has approximately equal surface energies. The results from [28] suggest this approach may yield films where the conventional Herring condition does indeed apply. Other methods to assess the effects of surface energy on TJ geometry could include an analysis where-in computed grain boundary energies are used to predict TJ geometries based on the Herring equation and deviations from this predicted microstructure are assessed as a function of the participating grains' surface energies. Other examples of experiments include eliminating substrate-induced residual strain by floating films off substrates before annealing and introducing variable encapsulation layers on the top and bottom surfaces, among many others.

The insight gained from the failure of the conventional Herring condition to describe these systems also points to future directions of theoretical study of the behavior of nanocrystalline thin films. A body of work simulating grain growth in thin films already exists; a number of mathematical models and simulations of thin-film grain growth apply the Herring condition *a priori* to govern local triple junction equilibrium with and without anisotropic grain boundary energy, see for example [44-46]; other approaches use finite triple junction mobilities [29,30,47]. Still more simulations have been developed to include the effects of strain energy, surface energy, impurity drag and grain boundary grooving on grain growth [40,48-51], but no model yet includes all of the necessary components to simultaneously reproduce the experimental observations of every geometric and topological metric of microstructure [52]. Based on the results presented here, naïve application of the Herring boundary condition is likely invalid in most cases relating to thin films. To understand boundary networks in these technologically important systems, it is critical that modelers and experimentalists work in tandem to systematically identify and incorporate the unique constraints of thin films, seeing as the simplest and most widely accepted model for junction geometry is empirically shown to fail in describing the behavior of the boundary network.

## 4. Conclusion

Orientation data collected via precession electron diffraction from a sputter deposited, nominally 40 nm-thick tungsten film and a sputter deposited, nominally 100 nm-thick aluminum film under three annealing conditions were analyzed to extract relative grain boundary energies under the assumption of Herring force balance at the triple junctions, using a recently reported technique. The GBCDs of the films were previously reported to correlate closely with comparable bulk materials, but the methodology for relative energy calculation, applied here, did not successfully reproduce expected energetic trends. The calculations did not recover a general inverse relationship between energy and population and did not produce GBEDs consistent with MD calculations. This result is in spite of clear correlations between the experimental populations and the theoretical GBEDs calculated from molecular dynamics and, in the case of tungsten, interpolation. This indicates that in this set of films, which includes two which have an opportunity to rearrange their boundary networks and two which have not, application of the Herring condition to solve for boundary energies is not appropriate, implying that the geometry of the triple junctions is not well-described by the Herring condition. This insight has important implications for simulations of grain growth in spatially constrained systems, where the grain size is larger than the thickness of the sample. Future work must include more detailed models of driving forces behind the migration of boundaries and the development of the network's junction geometry, as the commonly assumed model for triple junction geometry does not successfully describe the behavior of the network in these situations.

## Acknowledgements

MP and KB acknowledge support from the U.S. National Science Foundation (NSF) grant DMS-1905492 and the DMREF program under DMS-2118206. GR acknowledges support from the NSF under DMREF Grant DMR-2118945. SR and OC acknowledge the support of the Research Strengthening Project of the Faculty of Engineering, King Mongkut's University of Technology

Thonburi (KMUTT). ERH and GLWH acknowledge support from the NSF grant DMR-1817321. YE acknowledges partial support from NSF grant DMS-1905463 and the DMREF program under DMS-2118172.

**Figure Captions:**

Fig. 1. (a) Inverse pole figure map along the sample normal, showing orientations of grains in a representative field of view of the nominally 40 nm-thick tungsten film. Reconstructed boundary segments are drawn as thin black lines. (b) Inverse pole figure, along the sample normal, for this field of view; it shows very weak orientation texture, with a maximum of 1.33 multiples of random distribution at the [111] direction.

Fig. 2. Probability density of boundary disorientation angle across all reconstructed boundaries in all fields of view for the 40 nm-thick tungsten film plotted as a solid black line and the random, i.e. Mackenzie, distribution for disorientation [32] plotted as a dotted green line.

Fig. 3. Grain boundary plane distributions for the 40 nm-thick tungsten film, calculated using the stereological method [3] for Σ11, Σ17b, and Σ33a misorientations (a-c). The corresponding relative energy distributions are shown in the second column (d-f). Here, r.u. stands for relative units, and is referenced to the average relative boundary energy. Finally, in the third column (g-i), the energies calculated based on the interpolation scheme presented in [25] are shown.

Fig. 4. Logarithm of boundary populations, log(λ) with λ as relative area in units of MRD, binned by their relative energy for the 40 nm-thick tungsten film for misorientations of Σ3 (black circles), Σ11 (red squares), Σ17b (green triangles), and Σ33a (blue diamonds). Least squares fit lines are plotted in the same color as their corresponding misorientation.

Fig. 5. Representative inverse pole figure maps for the 100 nm-thick aluminum films in their (a) as-deposited state and the films after annealing at 400°C for (b) 30 minutes and (c) 150 minutes. The corresponding grain boundary plane distributions at a fixed misorientation of 60°|[111] (after twin adjustment) are shown below their maps in (c-f).

Fig. 6. Disorientation distribution for the 100 nm-thick aluminum films, reflecting the cleanup procedure used in the current work, plotted in solid black (as-deposited), red (30 minutes, 400°C), and blue (150 minutes, 400°C). The Mackenzie distribution for disorientations of randomly orientated cubes is shown as a dotted green line.

Fig. 7. Grain boundary plane distributions for the as-deposited aluminum film calculated using the stereological method and plotted in stereographic projection along the [001] direction for fixed misorientations of $\Sigma 3$, $\Sigma 7$, and $\Sigma 11$ (a-c). There are 4,491, 1,585, and 2,020 boundary segments represented in each figure respectively. The corresponding relative energy distributions are shown for the same set of boundaries in the second column (d-f). Here, r.u. stands for relative units, referenced to the average relative boundary energy. Finally, the absolute energies computed via molecular dynamics simulations are plotted in the third column (g-i). Note the minima at the (111) twist position for the $\Sigma 3$ and $\Sigma 7$ misorientations, with minimum values of 75 mJ/m$^2$ and 271 mJ/m$^2$ respectively, which are vanishingly narrow in the plots due to the limited sampling of the 5D space by the MD calculations [36,37]. Further, the appearance of the $\Sigma 7$ symmetry is affected by the contouring between sparse points during plotting.

Fig. 8. Grain boundary plane distributions for the aluminum film annealed at 400°C for 150 minutes calculated using the stereological method and plotted in stereographic projection along

the [001] direction for fixed misorientations of Σ3, Σ7, and Σ11 (a-c). There are 2,486, 760, and 1,158 boundary segments represented in each figure respectively. The corresponding relative energy distributions are shown for the same set of boundaries in the second column (d-f). Here, r.u. stands for relative units, and are referenced to the average relative boundary energy. Finally, the absolute energies computed via molecular dynamics simulations are plotted in the third column (g-i). Note the minima at the (111) twist position for the Σ3 and Σ7 misorientations, with minimum values of 75 mJ/m$^2$ and 271 mJ/m$^2$ respectively, which are vanishingly narrow in the plots due to the limited sampling of the 5D space by the MD calculations [36,37]. Further, the appearance of the Σ7 symmetry is affected by the contouring between sparse points during plotting.

Fig. 9. Logarithm of boundary populations, log(λ) with λ as relative area in units of MRD, as binned by their relative energy for the aluminum film at Σ3, Σ7, Σ11, and Σ19a misorientations for the (a) as-deposited film and (b) the film annealed for 150 minutes at 400°C. Least squares fit lines are plotted in the same color as their corresponding misorientation, including two regimes for the Σ3 boundaries.

Fig. 10. Logarithm of boundary populations, log(λ) with λ as relative area in units of MRD, in the as-deposited aluminum film plotted against boundary energies computed by molecular dynamics in [36,37]. All 7,304 examined boundaries are shown in (a) and (c), with the boundaries with misorientations specifically discussed in this work highlighted in color in (a) and (b), and boundary types with at least one boundary type with population >100 MRD

highlighted in (c) and (d). Least squares fit lines are plotted in the same color as their corresponding misorientation.

Fig. 11. Relative energies extracted from triple junction geometry of the as-deposited film for the 7,304 boundary types examined in [36,37], plotted against the grain boundary energies calculated via MD in [36,37]. The least squares best fit line is plotted in red, with a correlation coefficient < 0.1.

**Supplemental Information**

**Section S1: Cleanup**

The orientation data derived from precession enhanced electron diffraction must be cleaned up to be usable. The pseudosymmetry problem, in particular, makes distinguishing certain orientations related by well defined symmetry operators very difficult in spot patterns. For this reason, in grain size calculations, boundaries with misorientations corresponding to these orientations are generally removed [10,11,31] In the original GBCD analysis, to preserve as many boundaries as possible, points with confidence indices below a certain threshold were excluded [11]. Pseudosymmetry boundaries usually appear as zig-zags and tangles of reconstructed boundaries. In the work reported here, the pseudosymmetry cleanup was applied in the TSL OIM™ to remove certain misorientations from the data to minimize the number of spurious boundaries, in order to only include triple points which represented geometrically accurate junctions. Figure S1 shows a comparison of one field of view from the aluminum films. Here, it is evident that the majority of boundaries removed were spurious tangles, not real boundaries. Even for pseudosymmetry misorientations which are symmetrically equivalent to CSL-type boundaries, the vast majority are preserved after the cleanup.

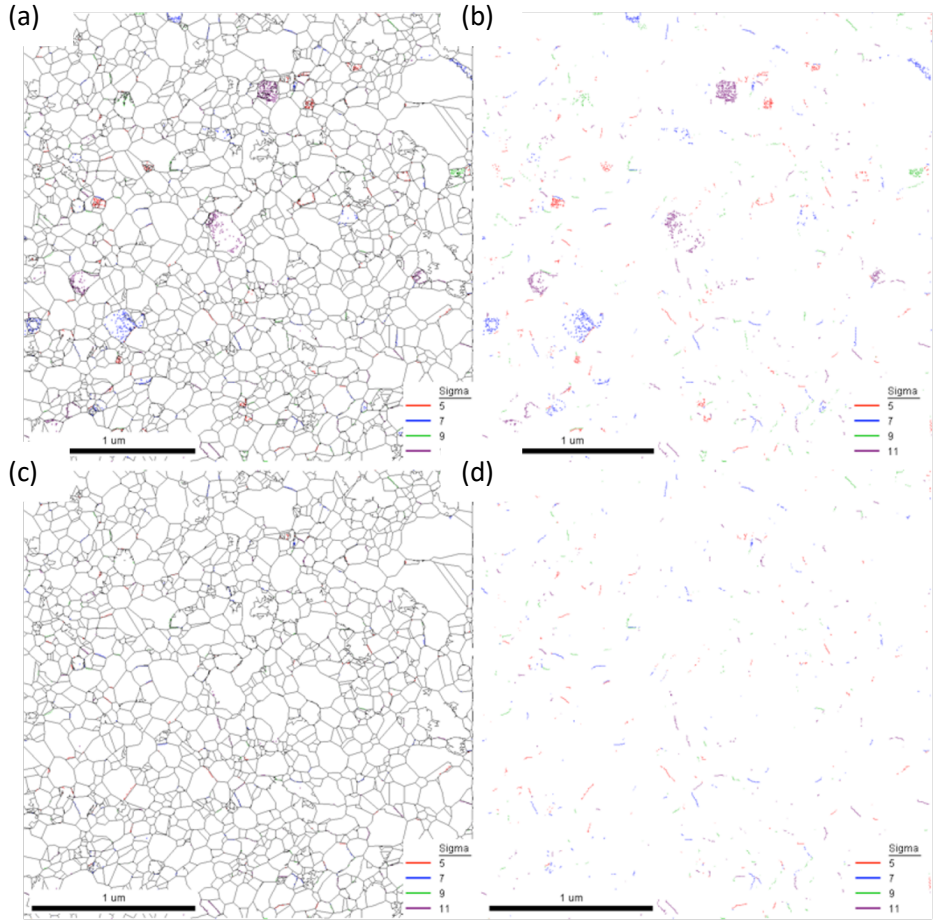

Figure S1: Comparison of locations of CSL misoriented boundaries identified by TSL OIM™, highlighted in color, with reconstructed boundary segments (a,c) and without reconstructed segments (c), before (a,b) and after (c,d) the pseudosymmetry cleanup procedure. Red, blue, green, and purple lines represent boundaries between points with misorientations corresponding to Σ5, Σ7, Σ9, and Σ11 boundaries, respectively.

This cleanup also had a notable impact on the disorientation distribution, as can be seen in Figure S2. The data-sets cleaned using the confidence index approach closely match the results presented in [11]. Applying the pseudosymmetry cleanup, as in this work, generally smooths the distribution and eliminates certain peaks in the distributions, notably near 39° and 50°, which were originally attributed to Σ9 and Σ11 boundaries respectively. As was seen in Figure S1, the apparent majority of removed boundaries with these misorientations were spurious. The smoother distribution, derived from the updated cleaning methodology more closely matches the Mackenzie distribution [32]. This change in cleanup also explains the difference in the exact MRD values observed between the GBCDs presented in this work and in [11].

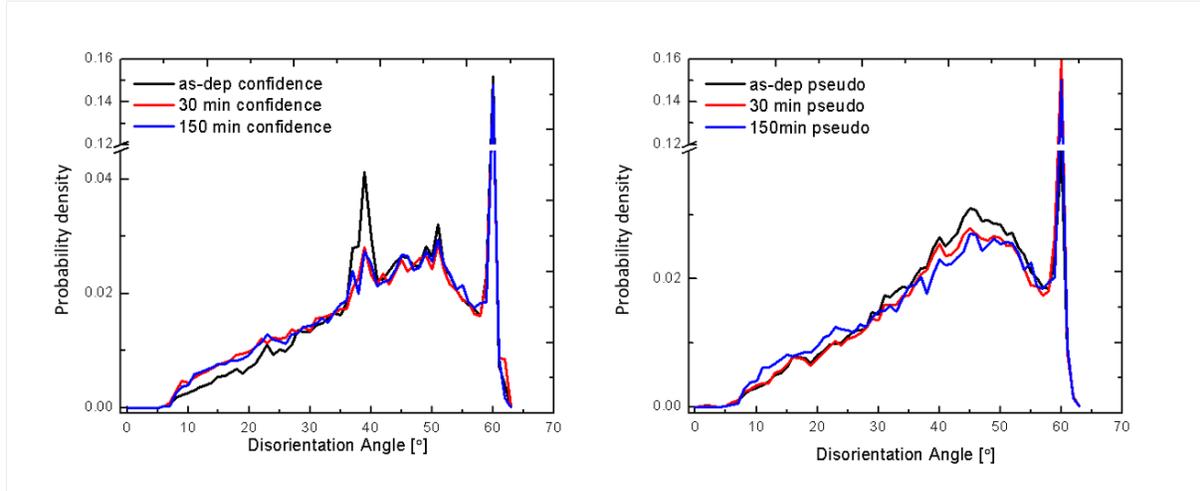

Figure S2: (a) Disorientation distribution generated using confidence-index cleanup as in [11], showing a distinct peak near disorientations of 40° and 50°. (b) Disorientation distribution generated using pseudosymmetry cleanup.

**Section S2: Columnar GBCDs and Twin Adjustment**

During this analysis, the adaptation of the energy extraction method relies on the columnarity of the grains and the implicit perpendicularity of the boundaries to the surfaces of the film. This is a commonly observed structure, and has been documented in thin films with grain sizes of the same order of magnitude or greater in size than the thickness of the film, including via cross sectional microscopy. As evidence, in these films, the grain boundary plane distribution (GBPD) is generally insensitive to whether the boundaries are treated as perpendicular to the film surface (and thus having plane normals parallel to the surface) or if their plane normals are determined stereologically. Figures S3, S4 compare the GBPDs calculated for the tungsten and as-deposited aluminum film from stereology and the columnar assumption. In both cases, while the exact intensities and spread of the peaks vary, the distributions show features in the same crystallographic locations, with similar relative intensities.

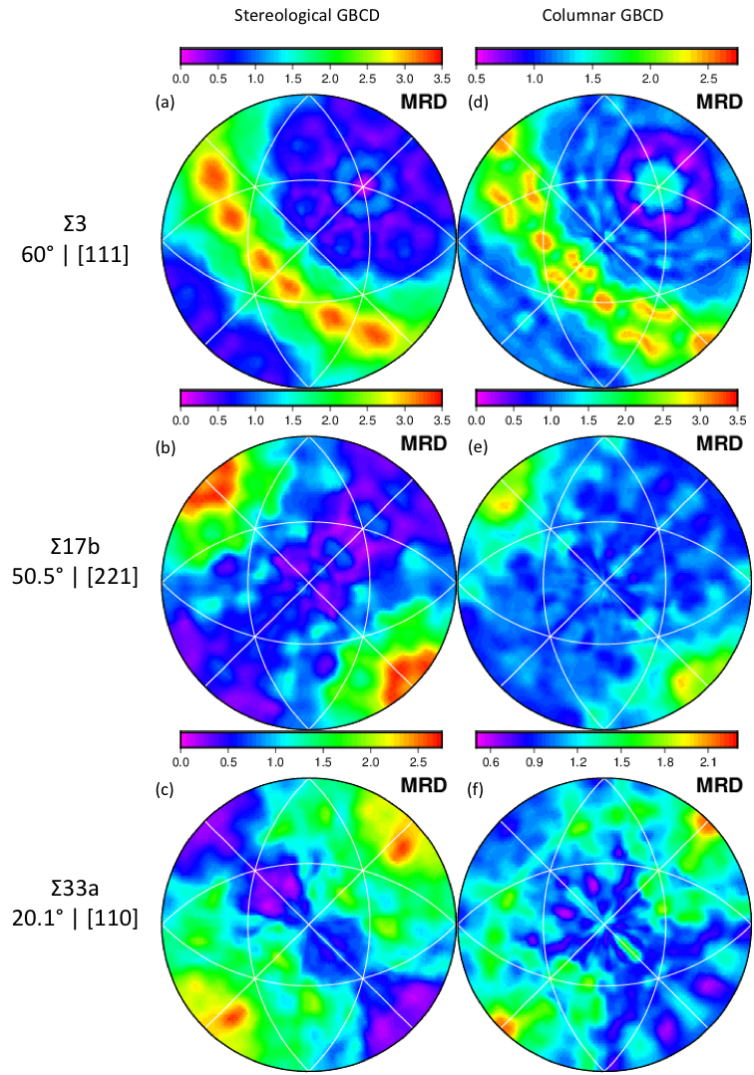

Figure S3: Grain boundary plane distributions for the Σ3, Σ17b, and Σ33a misorientations in the tungsten data-set. These include: (a-c) grain boundary plane distributions generated by the stereological method and (d-f) grain boundary plane distributions generated using a structured assumed to be completely columnar.

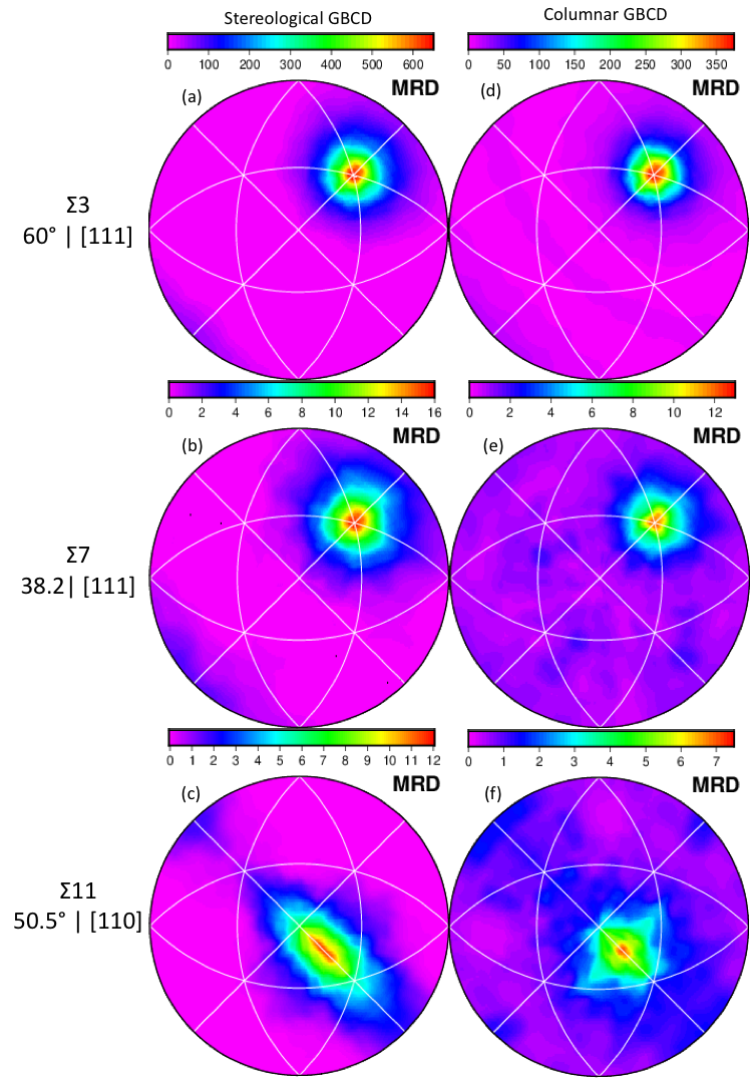

Figure S4: Grain boundary plane distributions for the Σ3, Σ7, and Σ11 misorientations in the as-deposited aluminum data-set. These include: (a-c) grain boundary plane distributions generated by the stereological method and (d-f) grain boundary plane distributions generated using a structure assumed to be completely columnar.

For the aluminum films, the grain boundary triple lines were modified to account for some boundaries which were known to break the general columnar rule. Figure S5 shows the GBPD for the 150-minute annealed aluminum film calculated stereologically and using a purely columnar assumption. It also shows the effects of the twin adjustment procedure on relative energies obtained from the extraction method. Without this adjustment, no expected features are recovered, as is evident in Figure S5 (g-i). After the adjustment, while the inverse correlation with population is still not observed for most boundaries, minima at the (111) twist positions and maxima at the (111) tilt positions become especially apparent for misorientations about [111].

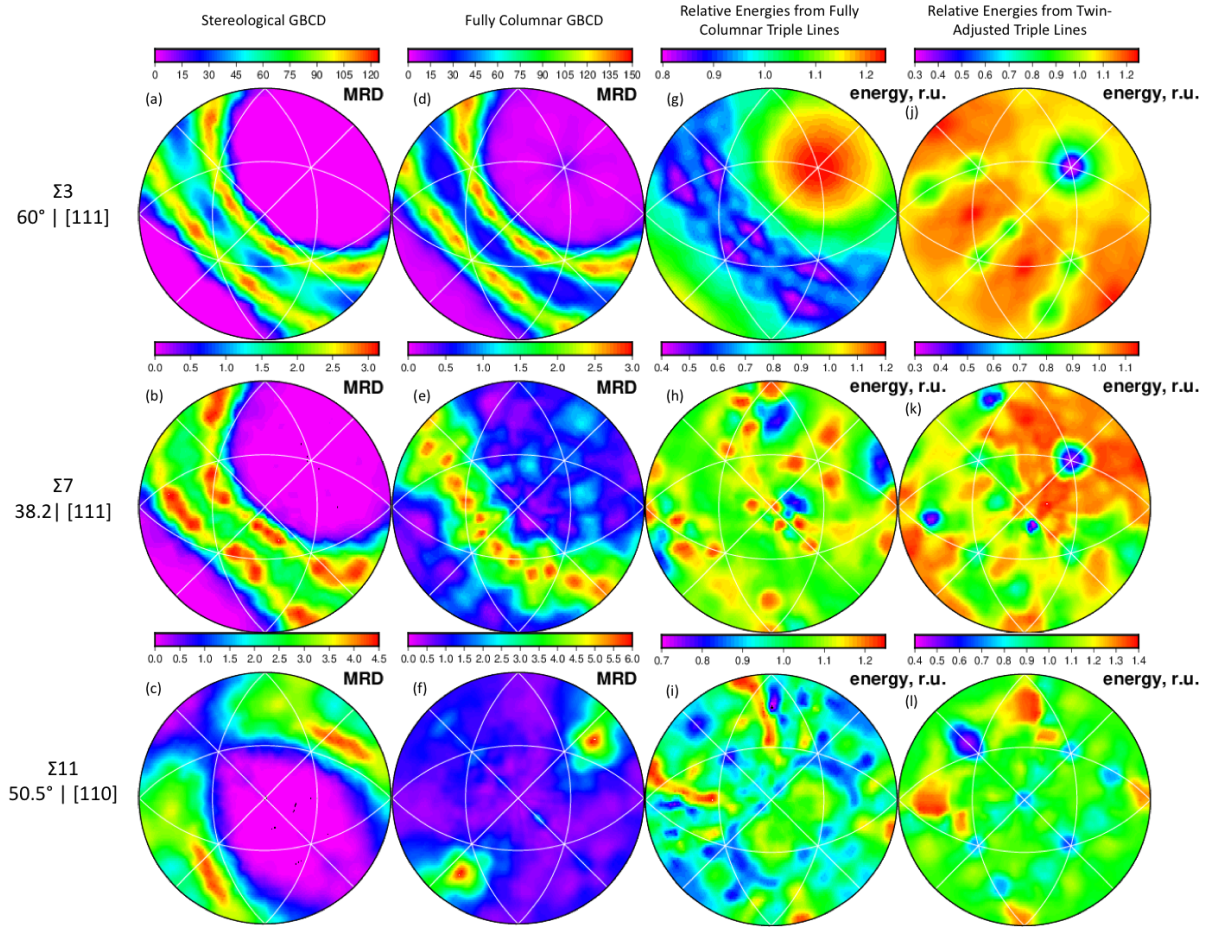

Figure S5: Grain boundary plane distributions for the Σ3, Σ7, and Σ11 misorientations in the as-deposited aluminum data-set. These include: (a-c) grain boundary plane distributions generated by the stereological method and (d-f) grain boundary plane distributions generated using a structure assumed to be completely columnar, and (g-i) the extracted relative energies of the grain boundaries without adjusting twin boundaries, and finally (j-l) the relative grain boundary energies extracted after twin adjustment.

## Section S3: Aluminum Film Annealed for 30 Minutes

The results presented in the main body of this work did not include those obtained from the 30 minute film; relative energies from this film are remarkably similar to those presented for the as-deposited film and the film annealed for 150 minutes at 400°C. Figure S6 presents the GBPD and relative GBED extracted after twin adjustment from the aluminum film annealed for 30 minutes; they are analogous to Figures 7 and 8 in the main body.

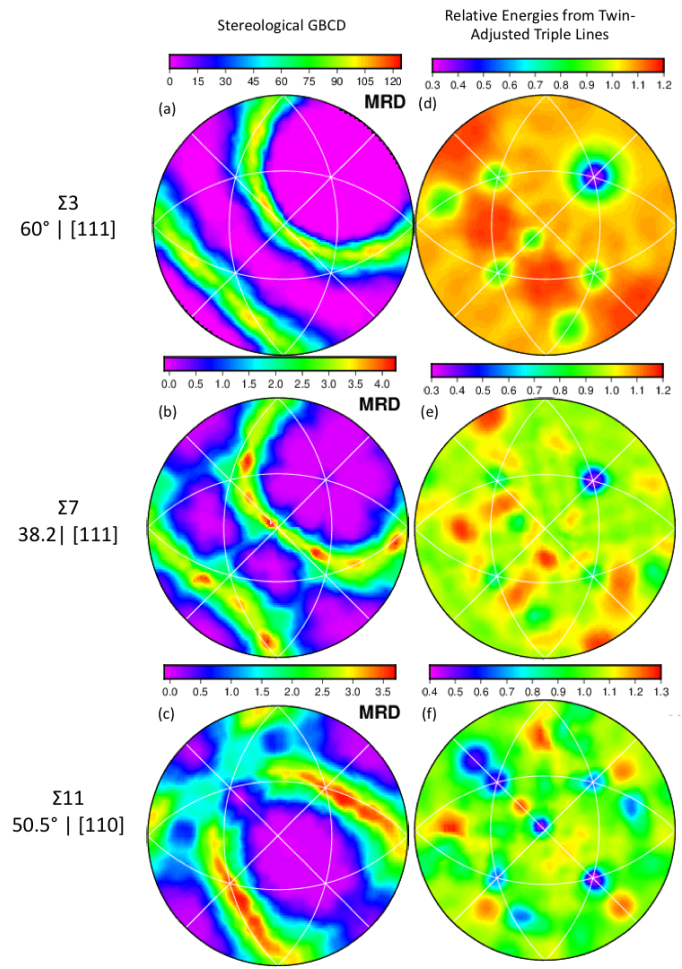

Figure S6: Grain boundary plane distributions for the Σ3, Σ7, and Σ11 misorientations in the aluminum annealed for 30 minutes at 400°C. These include: (a-c) grain boundary plane distributions generated by the stereological method and (d-f) grain boundary relative energy distributions generated using a structure assumed to be completely columnar.

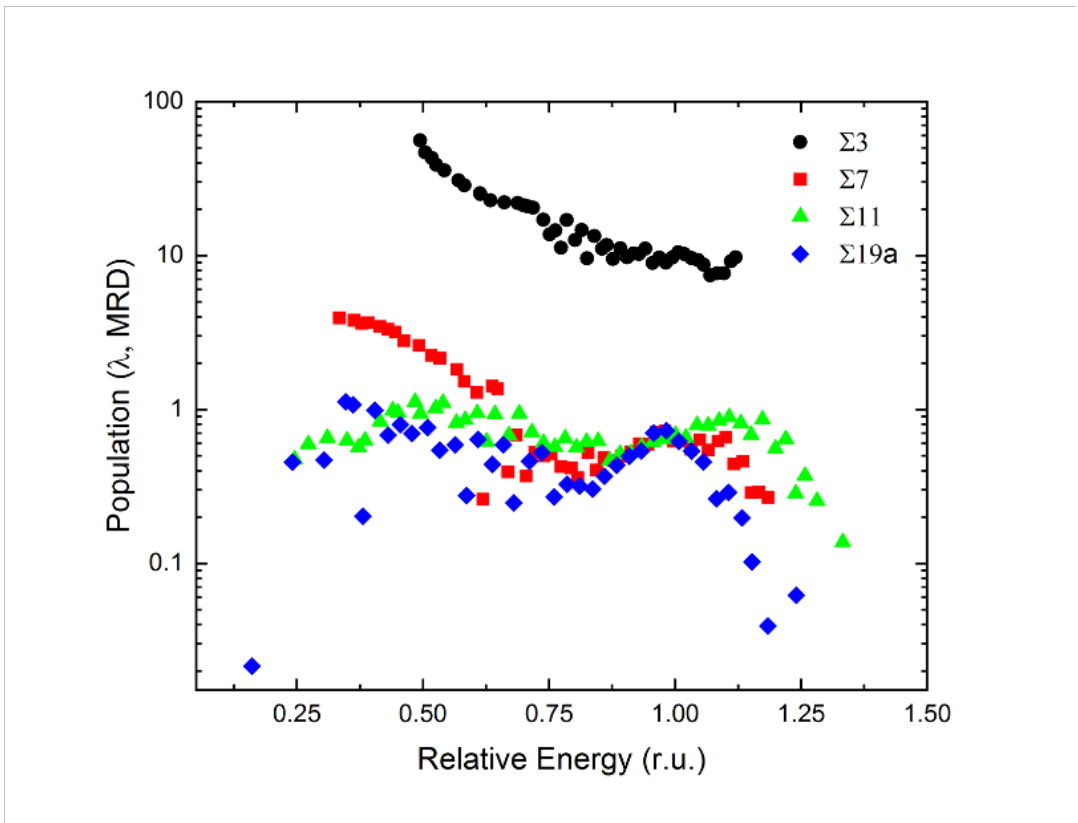

Figure S7: Boundary populations as binned by their relative energy for the aluminum film annealed for 30 minutes at 400°C at Σ3, Σ7, Σ11, and Σ19a misorientations.